\documentclass[aps,prl,twocolumn,groupedaddress,showpacs]{revtex4}
\usepackage{bm}
\usepackage{graphicx}
\usepackage{color}

\begin{document}

\title{Direct Observation of the Topological Surface States in Lead-Based Ternary Telluride Pb(Bi$_{1-x}$Sb$_x$)$_2$Te$_4$}
\author{S. Souma$^1$, K. Eto$^2$, M. Nomura$^3$, K. Nakayama$^3$,\\
T. Sato$^3$, T. Takahashi$^{1,3}$, Kouji Segawa$^2$, and Yoichi Ando$^2$}
\affiliation{$^1$WPI Research Center, Advanced Institute for Materials Research, 
Tohoku University, Sendai 980-8577, Japan}
\affiliation{$^2$Institute of Scientific and Industrial Research, Osaka University, Ibaraki, Osaka 567-0047, Japan}
\affiliation{$^3$Department of Physics, Tohoku University, Sendai 980-8578, Japan}

\date{\today}

\begin{abstract}
We have performed angle-resolved photoemission spectroscopy on Pb(Bi$_{1-x}$Sb$_x$)$_2$Te$_4$, which is a member of lead-based ternary tellurides and has been theoretically proposed as a candidate for a new class of three-dimensional topological insulators (TIs).  In PbBi$_2$Te$_4$, we found a topological surface state with a hexagonally deformed Dirac-cone band dispersion, indicating that this material is a strong TI with a single topological surface state at the Brillouin-zone center.  Partial replacement of Bi with Sb causes a marked change in the Dirac carrier concentration, leading to the sign change of Dirac carriers from $n$-type to $p$-type. The Pb(Bi$_{1-x}$Sb$_x$)$_2$Te$_4$ system with tunable Dirac carriers thus provides a new platform for investigating exotic topological phenomena.

\end{abstract}

\maketitle

     Three-dimensional (3D) topological insulators (TIs) are a novel state of quantum matter where the insulating bulk hosts gapless topological surface states (SS) with a Dirac-cone energy dispersion \cite{HasanReview, SCZhangReview} which appear within the bulk excitation gap generated by the large spin-orbit coupling (SOC).  The SOC is also responsible for the helical spin texture, where the spin direction is locked perpendicularly to the momentum. This peculiar spin texture \cite{XiaNP, SoumaPRL} is the source of the dissipationless spin current that exists in equilibrium \cite{KaneScience}, as well as the immunity of Dirac fermions to the backward scattering \cite{HasanReview}. Moreover, it has been predicted \cite{FuKanePRL} that if superconductivity is induced in such spin-helical Dirac fermions, non-Abelian Majorana fermions that are the essential ingredient of topological quantum computing may emerge in the vortices.  Angle-resolved photoemission spectroscopy (ARPES) has played a crucial role in the discovery of 3D TIs owing to the momentum ({\bf k}) resolving capability and the high surface sensitivity.  In fact, previous ARPES studies have identified a few types of new 3D TIs in the compounds containing group-V elements bismuth (Bi) and/or antimony (Sb), by observing the complex topological SS in Bi$_{1-x}$Sb$_x$ alloys \cite{HasanBiSbNature, HasanBiSbScience, MatsudaPRB} and the simpler topological SS with a single Dirac cone in tetradymite semiconductors (Bi$_2$Se$_3$ and Bi$_2$Te$_3$) \cite{XiaNP, Shen23} and thallium-based ternary chalcogenides (TlBiSe$_2$ and TlBiTe$_2$) \cite{SatoPRL, KurodaTBSPRL, ShenTBSPRL}.
     
    Recently, it has been proposed on the basis of the band structure calculations \cite{Xu_condmat, calc_Jin, calc_Mensh} that lead (Pb)-based layered ternary chalcogenides PbM$_2$X$_4$ (M: Bi, Sb, X: Se, Te) are a promising candidate for a new class of 3D TIs.  According to the calculations, some compounds of this system show a single Dirac cone at the Brillouin-zone (BZ) center originating form the topological SS, and a band inversion takes place at the Z or $\Gamma$ point in the bulk BZ \cite{calc_Mensh, calc_Jin} as suggested by the parity analysis of valence-band wave functions \cite{calc_Mensh}.  However, a clear experimental demonstration of the TI nature of Pb-based ternary chalcogenides has not yet been made, mainly due to the lack of high-quality single crystals.  Finding a new class of 3D TIs with possibly tunable Dirac carrier properties is particularly important, since only a few TI materials have been found to date and it has been difficult to tune the carrier type of the Dirac fermions in bulk TI crystals.
    
    In this Letter, we report a high-resolution ARPES study of Pb-based ternary chalcogenides Pb(Bi$_{1-x}$Sb$_x$)$_2$Te$_4$.  We have experimentally established that this system constitutes a new family of 3D TIs, by observing a single Dirac-cone-like surface band structure at the BZ center.  We also demonstrate that the chemical potential $\mu$ (and thus the sign of the Dirac carriers) is tunable by controlling the Bi/Sb ratio.  We discuss the present ARPES results in relation to the band calculations as well as to previous ARPES studies of other TIs.

    Single crystals of Pb(Bi$_{1-x}$Sb$_x$)$_2$Te$_4$ with $x$ = 0.0 and 0.4 were grown by melting stoichiometric amounts of high-purity elements (Pb 99.998$\%$, Bi, Sb, and Te 99.9999$\%$) in a sealed evacuated quartz tube at 900$^{\circ}$C for 1d, followed by slow cooling to 500$^{\circ}$C in about 4d and then switching off the furnace. On the other hand, the $x$ = 1.0 crystal is difficult to grow from the stoichiometric melt \cite{Shelimova}; in the present work, we melted the high-purity elements Pb, Sb, and Te with the molar ratio of 2:6:11 in a sealed evacuated quartz tube at 900$^{\circ}$C for 1d and then cooled it to room temperature at a rate of 48$^{\circ}$C/h. The PbSb$_2$Te$_4$ crystals were found at the bottom of the solidified rod and its crystal structure was confirmed by x-ray diffraction. Pb(Bi$_{1-x}$Sb$_x$)$_2$Te$_4$ has a rhombohedral crystal structure with the layers stacked along the [111] direction as shown in Fig. 1(a).  It consists of septuple layers where Pb atoms are sandwiched by Te-(Bi/Sb)-Te layers, with a real-space inversion symmetry centered at the Pb atom.  Cleaving of the crystal takes place between two septuple layers which are weakly coupled by van der Waals interactions.  The bulk BZ and its projected surface BZ onto the (111) plane are shown in Fig. 1(b), where the bulk BZ is shrunk along the [111] direction by $\sim$30$\%$ with respect to that of the prototypical TI Bi$_2$Te$_3$ owing to the longer $c$-axis.

\begin{figure}[t]
\includegraphics[width=3.4 in]{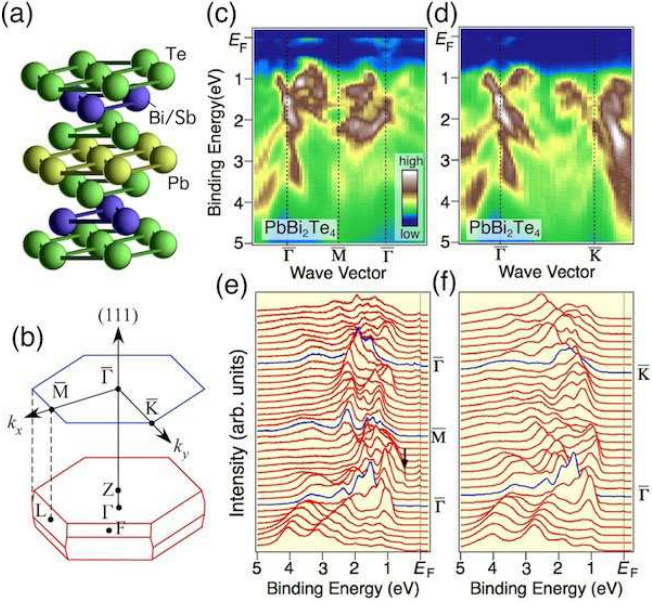}
\vspace{-0.4cm}
\caption{(Color online) (a) Crystal structure of Pb(Bi$_{1-x}$Sb$_x$)$_2$Te$_4$.  (b) Bulk BZ (red) and corresponding (111) surface BZ (blue).  (c), (d) ARPES intensity of PbBi$_2$Te$_4$ ($x$ = 0.0) plotted as a function of $E_{\rm B}$ and wave vector along $\bar{\Gamma}$$\bar{M}$and $\bar{\Gamma}$$\bar{K}$ high-symmetry lines, respectively, measured with $h\nu$ = 80 eV.  (e), (f) Corresponding energy distribution curves (EDCs) of (c) and (d), respectively
}
\end{figure}

   ARPES measurements were performed with a VG-Scienta SES2002 electron analyzer with a tunable synchrotron light at the beamline BL28A at Photon Factory (KEK).  We used circularly polarized light of 36-80 eV.  The energy and angular resolutions were set at 15-30 meV and 0.2$^{\circ}$, respectively.  Samples were cleaved {\it in-situ} along the (111) crystal plane in an ultrahigh vacuum of 1$\times$10$^{-10}$ Torr.  The Fermi level ($E_{\rm F}$) of the samples was referenced to that of a gold film evaporated onto the sample holder.  All the spectra were recorded at $T$ = 30 K.
   
  Figures 1(c) and 1(d) show valence-band ARPES spectra of the end member PbBi$_2$Te$_4$ ($x$ = 0.0) measured along high-symmetry lines $\bar{\Gamma}$$\bar{M}$ and $\bar{\Gamma}$$\bar{K}$ in the surface BZ, respectively.  Corresponding energy distribution curves (EDCs) are also displayed in Figs. 1(e) and 1(f).  We immediately notice in Figs. 1(c) and 1(d) several dispersive bands at the binding energy ($E_{\rm B}$) of 0.5-4.5 eV.  These bands are attributed to the hybridized states of Bi/Pb 6$p$ and Te 5$p$ orbitals \cite{Xu_condmat, calc_Jin, calc_Mensh}.  As seen in the EDCs along the $\bar{\Gamma}$$\bar{M}$ direction in Fig. 1(e), we find a holelike band with the top of  dispersion slightly away from the $\bar{\Gamma}$ point (see black arrow) which corresponds to the top of the valence band (VB) with the dominant Te 5$p$ character, in agreement with the calculations which show a local maximum between Z and L points of the bulk BZ \cite{calc_Mensh, calc_Jin}.  Such a local maximum structure is not reproduced in the calculations without including SOC \cite{calc_Jin}, suggesting that the observed VB feature is a fingerprint of the strong SOC.  One can also see in Figs. 1(c) and 1(e) a weak intensity in the vicinity of $E_{\rm F}$ in the {\bf k} region between the $\bar{\Gamma}$ and $\bar{M}$ points, which is assigned to the bulk conduction band (CB) of the strong Bi/Pb 6$p$ character that lies above $E_{\rm F}$ in the calculations \cite{Xu_condmat, calc_Jin, calc_Mensh}.  In the present ARPES experiment, this band is shifted downward to cross $E_{\rm F}$, possibly owing to the electron-doped nature of as-grown crystals as in the case of other 3D TIs \cite{XiaNP, Shen23, SatoPRL, KurodaTBSPRL, ShenTBSPRL}.
  
 \begin{figure}[t]
\includegraphics[width=3.4in]{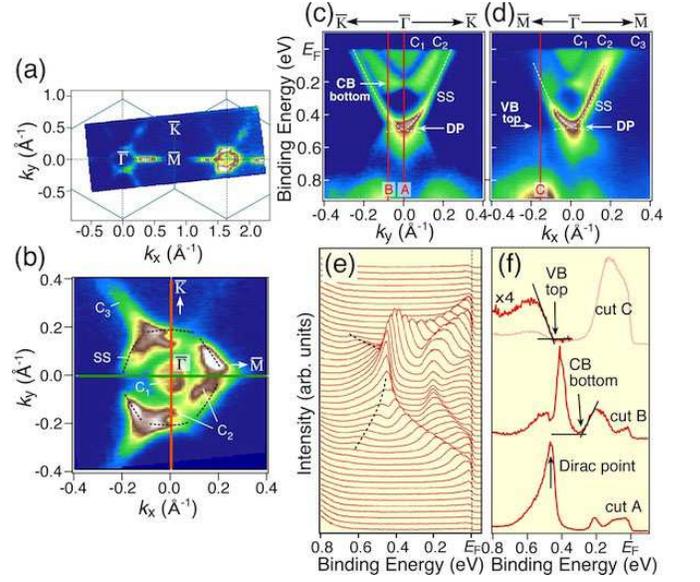}
\vspace{-0.4cm}
\caption{(Color online) (a) ARPES intensity plot at $E_{\rm F}$ of PbBi$_2$Te$_4$ ($x$ = 0.0) as a function of 2D wave vector measured with $h\nu$ = 80 eV.  (b) 2D intensity plot at $E_{\rm F}$ around the $\bar{\Gamma}$ point measured with $h\nu$ = 60 eV.  (c), (d) Near-$E_{\rm F}$ ARPES intensity as a function of $E_{\rm B}$ and wave vector along $\bar{\Gamma}$$\bar{K}$ and $\bar{\Gamma}$$\bar{M}$ cuts, respectively [orange and green lines in (b)].  Dashed lines are a guide for eyes to trace the SS.  (e) Corresponding EDCs along the $\bar{\Gamma}$$\bar{K}$ direction.  (f) Plot of EDCs at {\bf k} points A-C as indicated by red lines in (c) and (d).
}
\end{figure}
  
    Figure 2(a) displays the ARPES intensity at $E_{\rm F}$ of PbBi$_2$Te$_4$ plotted as a function of 2D wave vector measured with $h\nu$  = 80 eV, which covers a wide {\bf k} region in the surface BZ.  In both the first and second BZs, we find a circular-like Fermi surface (FS) centered at the $\bar{\Gamma}$ point together with elongated intensity patterns extending toward the $\bar{M}$ point.  To elucidate the FS topology and band dispersion around the $\bar{\Gamma}$ point in more detail, we have performed ARPES measurement with higher energy resolution and finer {\bf k} interval with $h\nu$ = 60 eV, and the result of the FS mapping is shown in Fig. 2(b).  We also plot the band dispersions for the $\bar{\Gamma}$$\bar{K}$ and $\bar{\Gamma}$$\bar{M}$ directions in Figs. 2(c) and 2(d), respectively.  
 
 \begin{figure}[t]
\includegraphics[width=3.4in]{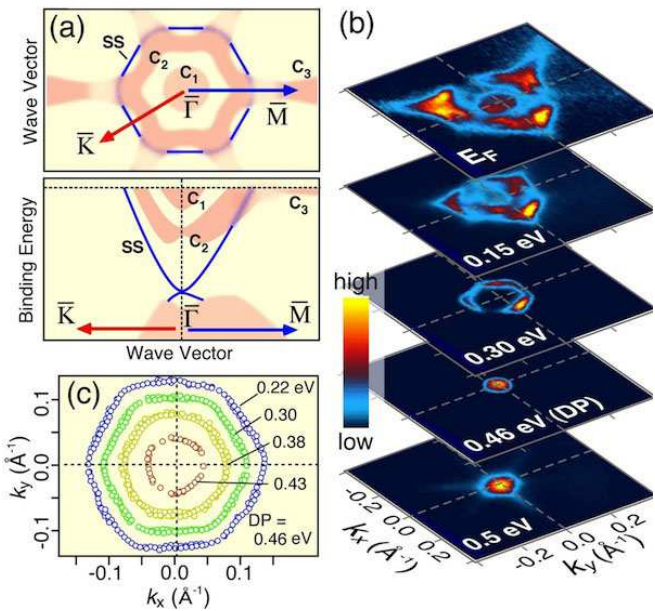}
\vspace{-0.4cm}
\caption{(Color online) (a) Schematic diagrams for (top) the FS and (bottom) the band dispersion of PbBi$_2$Te$_4$ obtained from the present ARPES experiment.  For the bulk CB and VB, a finite $k_z$ dispersion and momentum broadening have been taken into account.  (b) ARPES intensity plots for various $E_{\rm B}$'s between $E_{\rm F}$ and 0.5 eV.  (c) Energy contour plots of the SS for various $E_{\rm B}$'s estimated by tracing the peak position of momentum distribution curves (MDCs).
}
\end{figure}

\begin{figure*}[t]
\includegraphics[width=6in]{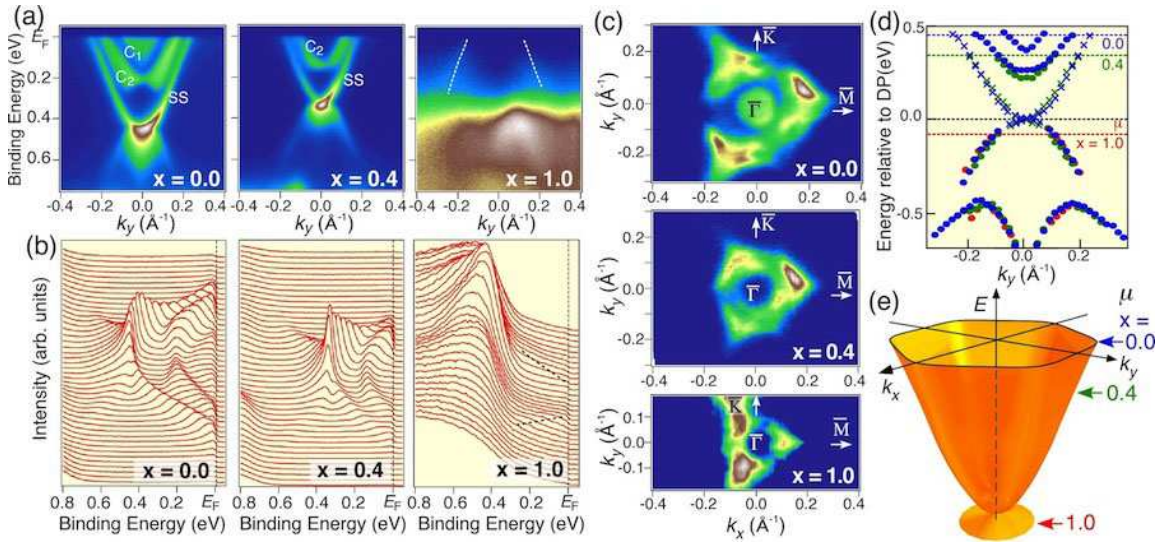}
\vspace{-0.4cm}
\caption{(Color online)  (a)-(c) Comparison of (a) the near-$E_{\rm F}$ band dispersions, (b) the EDCs, and (c) the FS in Pb(Bi$_{1-x}$Sb$_x$)$_2$Te$_4$ for three $x$ values ($x$ = 0.0, 0.4, and 1.0).  (d) The bulk (circles) and surface (crosses) band dispersions relative to the Dirac point for the three compositions x = 0.0, 0.4 and 1.0, estimated by tracing the peak positions of EDCs and MDCs. The locations of the chemical potential for the three compositions with respect to the Dirac point are shown by horizontal dashed lines. (e) Schematic 3D picture of the Dirac-band SS and the position of $\mu$ for each $x$.
}
\end{figure*}

 As clearly seen in the ARPES intensity of Fig. 2(c), we find a sharply dispersive ``V''-shaped electron-like band (denoted SS) whose bottom is located at $E_{\rm B}$ $\sim$ 0.5 eV.  A careful look at energy distribution curves (EDCs) in Fig. 2(e) further reveals the presence of an additional branch band showing holelike dispersion centered at the $\bar{\Gamma}$ point (traced by dashed lines). This branch merges into the V-shaped band exactly at the $\bar{\Gamma}$ point, satisfying the Kramers degeneracy and forming a deformed Dirac-cone dispersion as in Bi$_2$Te$_3$ \cite{Shen23}.  The surface origin of these branches was confirmed by the fact that the energy position is stationary with the variation of $h\nu$ (not shown).  Note that the energy position of other bands was also found to be insensitive to the $h\nu$ variation in the present range of $h\nu$, presumably because of the quasi-two-dimensional character of electronic states due to the long $c$-axis.  Nevertheless, the observed gapless dispersion should belong to the SS, since it does not show up in the bulk band calculations of PbBi$_2$Te$_4$ \cite{Xu_condmat, calc_Mensh}.  
 
 One can also see in Figs. 2(c) and 2(d) two electron-like bands within the SS near $E_{\rm F}$ which exhibit a parabola-like (C$_1$) and ``w''-shaped (C$_2$) dispersions, corresponding to a small circular pocket and a larger pocket with triangular-like intensity distribution, respectively, in Fig. 2(b) \cite{Matrixele}.  Furthermore, in Figs. 2(b) and 2(d) another weak intensity (C$_3$) is seen outside of the SS which either form an elongated pocket or are connected to the C$_2$ band along the $\bar{\Gamma}$$\bar{M}$ direction, although it is hard to experimentally conclude which is the case because of the presence of multiple bands in the close vicinity of $E_{\rm F}$ as well as the possible surface resonance which smears the overall intensity distribution in this {\bf k} region.  The C$_1$-C$_3$ bands are attributed to the bulk CB, and the bottom of the lowest-lying CB is located at the {\bf k} region where the C$_2$ band exhibits a local minimum structure, as indicated by a red line for cut B in Fig. 2(c).  Finally, in Fig. 2(d) the highest-lying bulk VB has its local maximum slightly away from the $\bar{\Gamma}$ point (at cut C), suggesting an indirect nature of the bulk band gap.
 
   We have quantitatively estimated the bulk band gap by determining the leading (trailing) edge of the VB (CB), as shown by an arrow in Fig. 2(f).  The energy location of the CB bottom and the VB top are $\sim$ 0.25 eV and $\sim$ 0.45 eV, respectively, corresponding to the band-gap size of $\sim$ 0.2 eV, slightly larger than that of the band calculations (0.1-0.15 eV) \cite{Xu_condmat, calc_Jin, calc_Mensh}.   Moreover, as seen in the EDC for cut A, the Dirac point (at $E_{\rm B}$ = 0.46 eV) is situated at almost the same energy as the VB top.  We have estimated the band velocity at the Dirac point, $v_{\rm Dirac}$, to be 1.0 eV${\rm \AA}$ along the $\bar{\Gamma}$$\bar{M}$ direction, which is only about 35-50$\%$ of the value obtained for Bi$_2$Te$_3$  and TlBiSe$_2$ (2.7 eV${\rm \AA}$ \cite{Shen23} and 2.0 eV${\rm \AA}$ \cite{SatoPRL}, respectively).
 
  We highlight in Fig. 3(a) the FS and band diagrams for the bulk and surface bands obtained from the present ARPES experiment.  We recognize complex and anisotropic nature of the CB which is distinctly different from prototypical TIs \cite{XiaNP, Shen23} showing a weaker anisotropy in the CB, likely because of the heavily electron-doped nature of PbBi$_2$Te$_4$ as apparent from the deeper energy position of the Dirac point as compared to other TIs \cite{XiaNP, Shen23, SatoPRL, KurodaTBSPRL, ShenTBSPRL}.  To show the 2D band dispersion, we plot in Fig. 3(b) the ARPES intensity for $x$ = 0.0 as a function of $k_x$ and $k_y$ for several $E_{\rm B}$'s.  The triangular-shaped intensity pattern at $E_{\rm F}$, which is dominated by the contribution from the C$_2$ band, gradually diminishes upon approaching the Dirac point.  At $E_{\rm B}$ = 0.3 eV, we recognize a strongly deformed ring-like intensity pattern which originates from the SS.  This ring-like image shrinks and converges into a single bright spot at the Dirac point ($E_{\rm B}$ = 0.46 eV), and then expands again below the Dirac point ($E_{\rm B}$ = 0.5 eV), as expected from the Dirac-cone energy dispersion.  At $E_{\rm B}$ = 0.5 eV, we also notice an additional weak intensity extending toward the $\bar{M}$ point, which originates from the top of the VB as also seen in Fig. 2(d).  To discuss the shape of the Dirac band in more detail, we have quantitatively estimated the {\bf k} location of the band crossing point in various $E_{\rm B}$ slices above the Dirac point, and a representative result is displayed in Fig. 3(c).  Obviously, the contour map of the band dispersion signifies the hexagonal warping which appears to become more prominent as one moves away from the Dirac point, similarly to the case of other TIs \cite{Shen23, Kuroda23PRL, SoumaPRL, SatoPRL}.
 
  Next we demonstrate the evolution of the electronic states upon Sb substitution for Bi in Pb(Bi$_{1-x}$Sb$_x$)$_2$Te$_4$.  Figures 4(a)-(c) display plots of near-$E_{\rm F}$ ARPES intensity around the $\bar{\Gamma}$ point as a function of $E_{\rm B}$ and  {\bf k}, corresponding EDCs, and the FS mapping, for $x$ = 0.0, 0.4, and 1.0, respectively.  The band dispersions for $x$ = 0.4 appear to be quite similar to that for $x$ = 0.0, although the overall dispersive features are shifted upward by $\sim$ 0.15 eV.  The C$_1$ band has completely disappeared in the ARPES intensity, as one can also confirm in the absence of the C$_1$ pocket in Fig. 4(c).  The C$_3$ FS also vanishes at $x$ = 0.4.  At $x$ = 1.0 (PbSb$_2$Te$_4$), on the other hand, the overall spectral feature appears to be quite different from those at $x$ = 0.0 and 0.4.  As seen in Fig. 4(a), we observe a holelike band with a weak intensity at $E_{\rm B}$ = 0-0.3 eV, together with a less dispersive bright intensity pattern at $\sim$ 0.6 eV.  The former band is assigned to the bulk VB which is seen at $E_{\rm B}$ = 0.4-0.6 eV at $x$ = 0.4, whereas the latter band is one of the VB which corresponds to the feature at $\sim$ 1.0 eV for $x$ = 0.0 [see Figs. 1(c)-(f)], suggesting that the Dirac point for $x$ = 1.0 is located in the unoccupied region.
  
     To quantitatively evaluate the evolution of the electronic states upon Sb substitution, we plot in Fig. 4(d) the band dispersions determined from the peak positions in the EDCs or MDCs.  One can recognize that the energy shift of the SS and bulk bands proceeds roughly in a rigid-band manner since the bands for different $x$ values nearly overlap with each other when we plot their energy positions with respect to the Dirac-point energy, despite the total $\mu$ shift of as large as 0.6 eV.   Indeed, the band dispersion below $E_{\rm F}$ for $x$ = 1.0 suggests that the Dirac point is located above $E_{\rm F}$ by $\sim$ 50 meV, pointing to the sign change of Dirac carriers from $n$-type to $p$-type at some $x$ between 0.4 and 1.0, in agreement with the Hall coefficient measured with our samples which exhibits a negative value at $x$ = 0.0-0.4 and a positive value at  $x$ = 1.0.  The present result thus demonstrates that the substitution of Bi with Sb results in the $\mu$ shift without significantly altering the shape of the Dirac SS, as highlighted in the schematic band picture in Fig. 4(e).
     
     The experimental realization of the sign change of Dirac carriers points to the high potential of Pb-based ternary chalcogenides for investigation of the various novel topological phenomena which requires the control of Dirac carrier conduction. It would also provide an excellent platform for the development of novel topological devices with $p$-$n$ junction configurations that are essential for various applications as in semiconductor technology as well as with dual-gate configuration for the electric control of spins \cite{Dual}.  Moreover, the observed variation in the chemical-potential value of $\sim$ 0.6 eV as achieved by the Bi/Sb replacement, which is the largest amongst known TI systems, is useful for varying the Dirac carrier concentrations in a wide range, though bulk carriers coexist in most of the ranges.  A next important step would be to synthesize a truly bulk insulating sample, which could be accessible by a fine control of the Bi/Sb ratio in the crystal.
     
     In summary, we have reported the ARPES study on Pb-based ternary tellurides Pb(Bi$_{1-x}$Sb$_x$)$_2$Te$_4$. We found direct evidence for the Dirac-cone topological SS within the indirect bulk band gap.  Moreover, we have revealed a sign change of Dirac carriers from $n$-type to $p$-type upon variation in the Sb concentration $x$, demonstrating tunable Dirac carriers.  The Pb-based ternary chalcogenide system is thus a promising candidate for investigating novel topological phenomena in TIs.

\begin{acknowledgments}
We thank Y. Tanaka, E. Ieki, K. Kosaka, K. Yoshimatsu, H. Kumigashira, and K. Ono for their assistance in ARPES measurements.  This work was supported by JSPS (NEXT Program, KAKENHI 23224010, and Grant-in-Aid for JSPS Fellows 23$\cdot$4376), JST-CREST, MEXT of Japan (Innovative Area ``Topological Quantum Phenomena''), AFOSR (AOARD 10-4103), and KEK-PF (Proposal number: 2010G507).\end{acknowledgments}


\begin{references}
\bibitem{Matrixele}
Triangular intensity distribution is likely caused by the photoionization matrix-element effect.
\end{references}
\end{document}